\documentclass{article}
\usepackage{spconf,amsmath,graphicx,hyperref,amsfonts}
\usepackage{booktabs}
\usepackage{multirow}
\usepackage{array}
\usepackage[table]{xcolor}
\usepackage[caption=false,font=footnotesize]{subfig}
\usepackage{cite}
\usepackage{url}
\usepackage{hyperref}
\usepackage[T1]{fontenc}
\usepackage[utf8]{inputenc}
\usepackage{microtype}

\title{REAL: Reward Modeling through Evidence-Linked Analysis of Lifelikeness in Generated Speech}

\name{\shortstack{Yaodong Song$^{1,2}$, Jie Lian$^2$, Xusheng Yang$^2$, Genliang Zhao$^2$, Hongjie Chen$^2$,\\
Jian Kang$^2$, Zhen-Hua Ling$^1$, Jie Li$^{2,*}$, Yongxiang Li$^{2,*}$}\thanks{$^*$ Corresponding authors.}}
\address{$^1$University of Science and Technology of China, Hefei\\
$^2$Xingchen AGI Lab, China Telecom Artificial Intelligence Technology Co., Ltd.}

\begin{document}
\ninept
\maketitle
\begin{abstract}
As modern text-to-speech systems become increasingly intelligible and natural, evaluation is shifting from detecting gross synthesis failures to identifying subtle departures from lifelike human delivery. Because large-scale listening tests are costly and slow, speech reward models can serve as scalable automatic judges for comparing generated utterances, curating data, and supporting iterative TTS development. Existing speech reward models can predict scores or pairwise preferences and may generate natural-language rationales, but their supervision is typically defined at the utterance or pair level. The connection between local acoustic phenomena and the final preference therefore remains implicit. We introduce \textbf{REAL}, a framework for \textbf{R}eward modeling through \textbf{E}vidence-linked \textbf{A}nalysis of \textbf{L}ifelikeness in generated speech. Its central idea is to supervise not only the final pairwise preference but also a structured account of its supporting observations, linking local audio spans to perceptual phenomena, affected dimensions, and the overall judgment. Specifically, REAL-Jury constructs these evidence-linked reports through timestamp alignment, multi-model analysis, and dispute-oriented deliberation. REAL-Judge learns to generate them from paired audio inputs. REAL-Eval assesses preference prediction in two tiers, from salient perceptual differences to subtle distinctions among strong TTS systems. Instantiated with MiMo-Audio, REAL-Judge improves pairwise preference accuracy over its backbone by 16.01 and 11.41 percentage points on the two tiers, respectively, and outperforms the open-source speech reward models SpeechJudge and UniSRM on both tiers.
\end{abstract}
\begin{keywords}
Speech Reward Model, Text-to-Speech Evaluation, Speech Language Model
\end{keywords}
\section{Introduction}
\label{sec:intro}

Recent advances in generative modeling have steadily narrowed the perceptual gap between text-to-speech (TTS) outputs and human recordings~\cite{cosyvoice2,maskgct,F5-TTS,goat_tts}. As synthesis quality continues to improve, determining whether generated speech is merely intelligible or genuinely resembles natural human delivery has become increasingly important for system comparison, data curation, and iterative model development. The prevailing mean opinion score (MOS) protocol~\cite{itu_p800}, however, depends on large-scale listening tests that are costly and time-consuming, and its outcomes can vary with listener composition and experimental conditions. More importantly, a single utterance-level score reveals neither where an utterance departs from natural human speech nor which acoustic phenomena drive the judgment. These limitations motivate scalable speech reward models that approximate human preferences while providing diagnostic signals beyond a scalar score.

Existing research on scalable speech assessment spans three main paradigms: scalar quality estimation, preference-based comparison, and generative evaluation. MOSNet~\cite{mosnet}, UTMOS~\cite{utmos}, and SCOREQ~\cite{scoreq} predict absolute perceptual quality, whereas UPPSQA~\cite{uppsqa} converts utterance-level scores into pairwise preferences. The emergence of audio language models~\cite{step_audio_r1,qwen2.5-omni,qwen3-omni} has further enabled reward models that produce judgments together with natural-language analyses. WavReward~\cite{wavreward} combines explicit reasoning with nonlinear reward learning, while SpeechJudge-GRM~\cite{speechjudge} integrates chain-of-thought supervised fine-tuning with reinforcement learning. SpeechLLM-as-Judges~\cite{speechllmjudges}, UniSRM~\cite{unisrm}, and TTS-PRISM~\cite{ttsprism} broaden automatic evaluation to multitask, multidimensional, and fine-grained diagnostic settings. Collectively, these advances have expanded the role of automatic evaluation from predicting how good an utterance sounds to comparing alternatives, diagnosing perceptual dimensions, and articulating the reasons behind a judgment.

However, richer outputs alone do not establish that the generated analyses are grounded in the audio evidence relevant to the preference decision. In existing data-construction pipelines, explanatory targets are often generated by a single speech-understanding model, while human utterance- or pair-level labels are used to retain samples whose final judgments agree with the annotated preferences. Such filtering constrains the correctness of the overall decision but does not independently verify the local acoustic claims contained in the explanation. Consequently, an explanation may survive preference-consistency filtering even when some of its supporting observations are not faithfully grounded in the audio. Prior studies~\cite{Kuan2024ObjectHallucination,SpotSound} have documented hallucinated audio events and limitations in fine-grained temporal grounding in speech-understanding models. More broadly, studies of chain-of-thought faithfulness~\cite{Turpin2023UnfaithfulCoT,Lanham2023CoTFaithfulness} show that generated reasoning can rationalize a conclusion post hoc rather than faithfully reflect the evidence used to reach it. Together, these findings indicate that agreement with an utterance-level preference label does not by itself guarantee an acoustically grounded explanation.

Motivated by these observations, we propose \textbf{REAL}, a framework for \textbf{R}eward modeling through \textbf{E}vidence-linked \textbf{A}nalysis of \textbf{L}ifelikeness in generated speech. The central design choice is to make the structured relationship between local observations and the overall preference an explicit supervision target. Specifically, an evidence-linked report identifies relevant audio spans, describes the associated perceptual phenomena, attributes them to affected dimensions, and explains their relevance to the pairwise judgment. These reports serve as training targets, so the model learns to generate temporally referenced comparative analyses together with its preference prediction. REAL implements this design through three components. \textbf{REAL-Jury} constructs the reports by coordinating independent speech-understanding models, aligning their observations in time, and revisiting disputed claims. \textbf{REAL-Judge} distills the resulting supervision into a single timestamp-aware reward model. \textbf{REAL-Eval} assesses preference prediction across two tiers, separating relatively salient perceptual differences from subtler distinctions among strong TTS systems. We will release the code and model weights to facilitate future research.

\section{METHOD}
\label{sec:method}

REAL uses an evidence-linked preference report as its supervision target, jointly representing the pairwise decision and temporally referenced observations that explain the comparison. It follows an evidence construction--distillation--evaluation pipeline. We detail the three stages of REAL below.

\subsection{REAL-Jury: Evidence Construction}

\begin{figure}[b]
    \centering
    \includegraphics[width=\columnwidth]{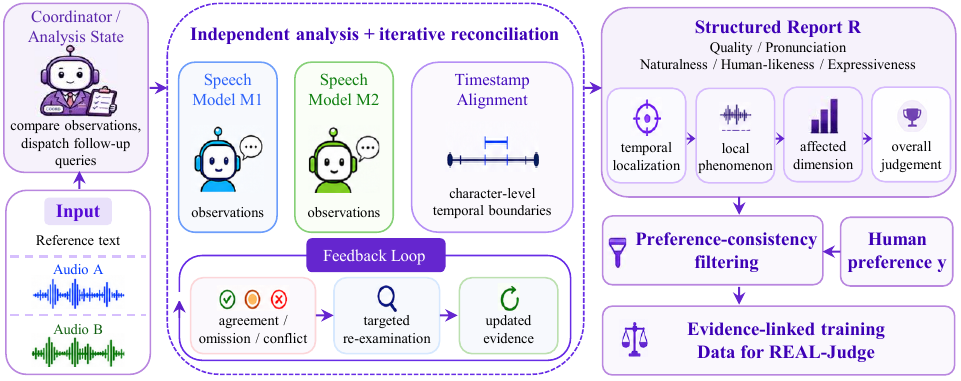}
    \caption{Overview of REAL-Jury. Independent analyses and timestamp alignment are iteratively reconciled into structured reports, which are filtered by human preferences to construct evidence-linked training data for REAL-Judge.}
    \label{fig:real-jury}
\end{figure}

Given a reference text $t$ and two candidate utterances $x^A$ and $x^B$, REAL-Jury constructs an evidence-linked preference report $R$ containing an overall prediction $\hat{y}$ and temporally localized observations describing where perceptually relevant differences occur, what speech phenomena they represent, and how they affect the final judgment, as illustrated in Fig.~\ref{fig:real-jury}.

\textbf{Jury composition.}
REAL-Jury adopts a modular and extensible orchestration architecture built around a central coordination model and a configurable pool of specialized models. In our implementation, the pool comprises two speech-understanding models and a timestamp model, while additional specialized audio-analysis models can be incorporated without changing the overall deliberation protocol. The coordination model maintains the analysis state, identifies unresolved questions, and dispatches targeted requests to the appropriate models. Conditioned on the reference text, each understanding model analyzes both candidates and reports perceptually salient differences. The timestamp model aligns each utterance with the reference text and returns character-level temporal boundaries, thereby establishing a shared temporal coordinate system.

\textbf{Dispute-oriented deliberation.}
Rather than aggregating one-shot model outputs, REAL-Jury operates as a closed-loop deliberation process. In the initial round, the coordination model requests independent analyses from both speech-understanding models without sharing their responses, while the timestamp model aligns each utterance with the reference text to establish shared temporal coordinates. The coordinator then compares the collected observations to identify agreements, omissions, and conflicts at the level of individual speech phenomena. When the models disagree about pronunciation, pausing, emotion, expressiveness, or the overall preference, it formulates targeted follow-up questions and asks the relevant models to re-examine specific characters or time spans. Newly collected evidence is incorporated into the analysis state before the next round. The process terminates when the outstanding discrepancies are reconciled or a predefined maximum number of rounds is reached, after which the coordinator consolidates the accumulated evidence into 
the preference report.

\textbf{Report organization and filtering.}
The final report forms a bottom-up chain from a time interval, through a local speech phenomenon and its affected dimension, to the final pairwise decision. Given an initial collection $\mathcal{D}$, samples whose independently predicted preferences disagree with the corresponding human labels are discarded after deliberation:
\begin{equation}
    \mathcal{D}_{\mathrm{filt}}
    = \left\{\left(t_i,x_i^{A},x_i^{B},R_i,y_i\right)
    \;\middle|\; \hat{y}_i = y_i,\; i=1,\ldots,|\mathcal{D}|\right\},
\end{equation}
where $R_i$ denotes the complete evidence-linked report produced by REAL-Jury.

\subsection{REAL-Judge: Evidence-Linked Reward Modeling}

\begin{figure}[t]
    \centering
    \includegraphics[width=\columnwidth]{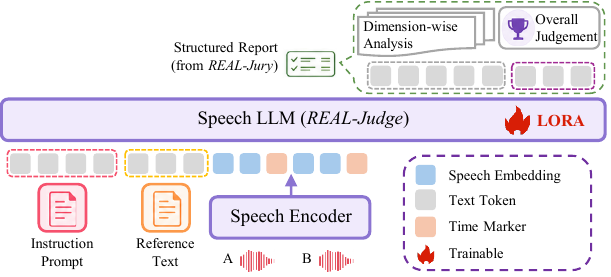}
    \caption{Training pipeline of REAL-Judge. A LoRA-adapted speech language model distills REAL-Jury's evidence-linked reports from instructions, reference text, and time-marker-augmented audio.}
    \label{fig:real-judge}
\end{figure}

REAL-Judge distills the supervision constructed by REAL-Jury into a single timestamp-aware reward model, as illustrated in Fig.~\ref{fig:real-judge}. Given an instruction $p$, reference text $t$, and candidates $x^A$ and $x^B$, it generates a structured report containing dimension-wise comparisons with temporally referenced observations, followed by an overall preference.

\textbf{Timestamp-aware representation.}
Fine-grained evidence localization requires the model to associate acoustic observations with explicit positions on the audio timeline. Relying only on the relative order of audio representations provides no direct temporal scale and can make event timing increasingly difficult to infer as the input becomes longer or more densely structured. To enable fine-grained temporal grounding, we follow recent audio and omni-modal language models that interleave explicit timestamps with audio representations~\cite{qwen3.5_omni,moss_audio}. Specifically, each candidate is encoded as $H^k=E_{\mathrm{aud}}(x^k)$, where $k\in\{A,B\}$. Explicit time-marker embeddings, denoted collectively by $M^k_{\Delta}$, are inserted at intervals of $\Delta$ seconds. The complete model input is organized as follows:
\begin{equation}
S=[p;t;\widetilde{H}^A;\widetilde{H}^B], \quad \widetilde{H}^k=\operatorname{Interleave}(H^k,M^k_{\Delta}).
\label{eq:time-marker}
\end{equation}

\textbf{Evidence-linked supervision.}
We perform supervised fine-tuning on $\mathcal{D}_{\mathrm{filt}}$ using low-rank adaptation (LoRA)~\cite{lora}, while freezing the backbone. For report tokens $R_i=(r_{i,1},\ldots,r_{i,L_i})$, the objective is the standard autoregressive negative log-likelihood
\begin{equation}
\mathcal{L}_{\mathrm{SFT}}
=-\sum_i\sum_{j=1}^{L_i}\log p_{\theta,\phi}
(r_{i,j}\mid S_i,r_{i,<j}),
\label{eq:sft}
\end{equation}
where $\theta$ and $\phi$ denote frozen backbone and trainable LoRA parameters, respectively. During inference, neither REAL-Jury nor human labels are required.

\subsection{REAL-Eval: Two-Tier Evaluation}
\label{sec:real-eval}

\begin{figure}[t]
    \centering
    \includegraphics[width=\columnwidth]{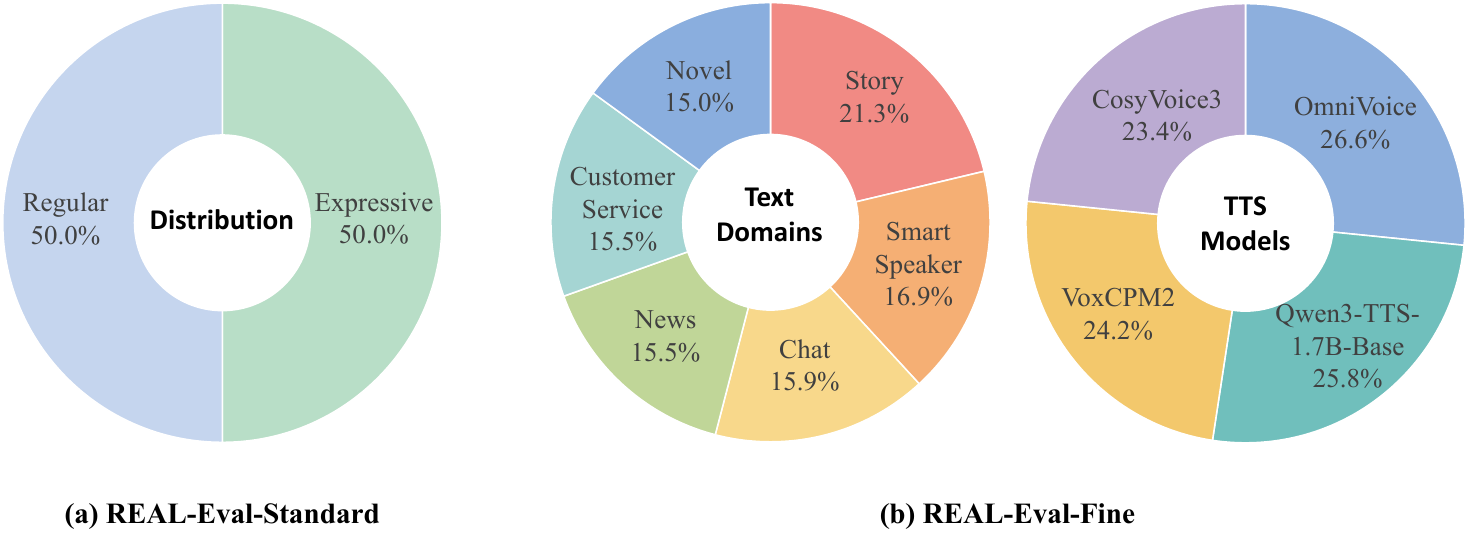}
    \caption{Composition of REAL-Eval. (a) REAL-Eval-Standard includes regular and expressive comparisons under cross-lingual and monolingual reference-to-target settings. (b) REAL-Eval-Fine covers six text domains and utterances generated by four strong TTS systems.}
    \label{fig:real-eval-data}
\vspace{-1.0em}
\end{figure}

REAL-Eval evaluates two complementary regimes rather than partitioning one distribution into easy and difficult samples. \textbf{REAL-Eval-Standard} targets comparisons containing relatively salient differences in basic quality or human-likeness, whereas \textbf{REAL-Eval-Fine} examines subtler delivery differences among strong contemporary TTS systems. In both tiers, evaluation is formulated as a binary comparison conditioned on a shared reference text, and the samples are post-processed to balance Audio A and Audio B preference labels. The data composition is summarized in Figure~\ref{fig:real-eval-data}.

\textbf{REAL-Eval-Standard.}
We retain 400 Chinese preference pairs from the regular and expressive subsets of SpeechJudge-Eval~\cite{speechjudge}. The candidate utterances are produced by six zero-shot TTS systems spanning autoregressive, flow-matching, and masked-generative architectures: ARS~\cite{maskgct}, MaskGCT~\cite{maskgct}, CosyVoice2~\cite{cosyvoice2}, CosyVoice2-INTP, Ints-INTP~\cite{intp}, and F5-TTS~\cite{F5-TTS}. The tier covers monolingual and cross-lingual reference conditions and tests whether a reward model recognizes basic departures in pronunciation completeness, acoustic stability, pausing, prosodic organization, and style-text consistency.

\textbf{REAL-Eval-Fine.}
We collect real-world texts from six application domains and apply text cleaning and normalization, including punctuation and number normalization and the removal of content unsuitable for speech synthesis. We then synthesize candidate utterances using OmniVoice~\cite{omnivoice}, Qwen3-TTS-1.7B-Base~\cite{qwen3_tts}, VoxCPM2~\cite{voxcpm2}, and CosyVoice3~\cite{cosyvoice3}. Ten annotators independently provide an overall preference judgment for each pair after jointly considering pronunciation accuracy, naturalness, human-likeness, and expressiveness. We discard pairs with $|A_{\mathrm{win}}-B_{\mathrm{win}}|\leq2$, retaining only decisions with at least a $7{:}3$ majority. The resulting 207 pairs capture delivery-related differences, such as whether pauses follow semantic structure, emphasis falls on informative words, rhythm and pacing remain natural, tone and emotion fit the text, and the delivery sounds mechanical, overly flat, or excessively dramatized.

\section{Experimental Setup}
\label{sec:experiments}

\subsection{Data and Configurations}

This work focuses on Chinese reference texts. The training corpus is constructed from selected preference pairs in the public SpeechJudge dataset~\cite{speechjudge}. Applying the complete REAL-Jury pipeline yields 11,947 pairs with evidence-linked supervision. DeepSeek-V4-Pro~\cite{deepseek_v4} serves as coordinator, orchestrating Qwen3.5-Omni~\cite{qwen3.5_omni} and Step-Audio-R1.1~\cite{step_audio_r1}. A multilingual Wav2Vec2 forced-alignment model\footnote{\url{https://dl.fbaipublicfiles.com/mms/torchaudio/ctc_alignment_mling_uroman/model.pt}} grounds their observations in time. The coordinator may issue up to 15 model calls per pair. REAL-Eval-Standard and REAL-Eval-Fine contain 400 and 207 pairs, respectively. The independently constructed REAL-Eval-Fine has no overlap with the training data in reference texts, reference speech, or source TTS systems.

\subsection{Training and Baselines}

We instantiate REAL-Judge with MiMo-Audio-Instruct~\cite{mimo_audio}. The LoRA rank and scaling factor are 64 and 128. Training uses a peak learning rate of $2\times10^{-4}$, cosine scheduling, a 0.03 warmup ratio, bfloat16 precision, and a global batch size of 32 on eight NVIDIA A800 80-GB GPUs. Time markers are inserted at a default interval of 2 seconds and aligned to valid boundaries in MiMo-Audio's packed multi-codebook token stream.

We compare against a lower-CER heuristic using the Mandarin Paraformer ASR model provided by the FunASR toolkit~\cite{funasr}; the closed-source models Gemini-3.1-Pro-Preview~\cite{gemini31pro} and Qwen3.5-Omni~\cite{qwen3.5_omni}; the open-source speech-understanding models Qwen3-Omni~\cite{qwen3-omni}, Step-Audio-R1.1~\cite{step_audio_r1}, and MiMo-Audio-Instruct~\cite{mimo_audio}; and the specialized reward models SpeechJudge~\cite{speechjudge} and UniSRM~\cite{unisrm}. Each system receives the reference text and two candidate utterances and predicts the human-preferred candidate. Micro-averages pool all samples within a tier, whereas macro-averages assign equal weight to its evaluation subsets.

\section{Results}
\label{sec:results}

\begin{table*}[th]
\centering
\caption{Preference accuracy (\%) on REAL-Eval-Standard and REAL-Eval-Fine. ``AVG'' is the macro-average over the corresponding subsets.}
\label{tab:baseline-comparison}
\resizebox{1.0\textwidth}{!}{%
\begin{tabular}{lccccccccccc}
\toprule
\multirow{2}{*}{\textbf{Model}} &
\multicolumn{3}{c}{\textbf{REAL-Eval-Standard}} &
\multicolumn{7}{c}{\textbf{REAL-Eval-Fine}} &
\multirow{2}{*}{\textbf{AVG}} \\
\cmidrule(lr){2-4}\cmidrule(lr){5-11}
& \textbf{Regular} & \textbf{Expressive} & \textbf{Micro}
& \textbf{Smart Speaker} & \textbf{Chat} & \textbf{News}
& \textbf{Customer Service} & \textbf{Story} & \textbf{Novel}
& \textbf{Micro} & \\
\midrule
Lower-CER & 20.50 & 17.00 & 18.75 & 5.71 & 9.09 & 3.12 & 9.38 & 6.82 & 6.45 & 6.76 & 9.76 \\
\midrule
Gemini-3.1-Pro-Preview & 82.33 & 77.33 & 79.83 & 53.34 & 67.68 & \textbf{72.92} & 61.46 & \textbf{75.00} & 72.04 & \underline{67.31} & 70.26 \\
Qwen3.5-Omni & 81.47 & 78.17 & 79.82 & \textbf{65.71} & \textbf{70.71} & \underline{71.88} & 62.50 & 59.09 & \textbf{79.57} & \textbf{67.63} & \textbf{71.14} \\
\midrule
Qwen3-Omni & 81.00 & 79.83 & 80.42 & 53.33 & 55.56 & 67.71 & 62.50 & 53.79 & 59.14 & 58.29 & 64.11 \\
Step-Audio-R1.1 & 74.00 & 76.67 & 75.33 & 54.29 & 62.63 & 54.16 & 61.46 & 55.31 & 59.14 & 57.65 & 62.21 \\
MiMo-Audio-Instruct & 70.83 & 66.00 & 68.42 & 49.52 & 57.58 & 50.00 & 56.25 & 51.52 & 60.22 & 53.95 & 57.74 \\
\midrule
SpeechJudge & 84.00 & \underline{82.00} & \underline{83.00} & 60.00 & 54.55 & 56.25 & \textbf{68.75} & 45.45 & 70.97 & 58.45 & 65.25 \\
UniSRM & \underline{84.50} & 77.00 & 80.75 & 57.14 & \underline{69.70} & 59.38 & 62.50 & \underline{63.64} & 67.74 & 63.29 & 67.70 \\
\midrule
\multicolumn{12}{l}{\textbf{Proposed Method}} \\
\rowcolor{gray!15}REAL-Judge & \textbf{85.50} & \textbf{83.36} & \textbf{84.43} & \underline{62.86} & \underline{69.70} & 67.86 & \underline{64.73} & 55.52 & \underline{75.58} & 65.36 & \underline{70.64} \\
\bottomrule
\end{tabular}}
\vspace{-1.0em}
\end{table*}

\subsection{Comparison with Baselines}

Table~\ref{tab:baseline-comparison} shows that the simple strategy of preferring the candidate with the lower CER agrees with human judgments on only 18.75\% of REAL-Eval-Standard pairs and 6.76\% of REAL-Eval-Fine pairs. To distinguish this pairwise comparison from the underlying transcription accuracy, we additionally compute the absolute CER over all candidate utterances, obtaining 6.56\% on REAL-Eval-Standard and 1.27\% on REAL-Eval-Fine. The low absolute CER on REAL-Eval-Fine indicates that the transcription errors are already rare. This is consistent with REAL-Eval-Fine's focus on delivery-related differences beyond transcription accuracy, that is, how speech is delivered rather than merely what is said.

REAL-Judge improves its MiMo-Audio backbone from 68.42\% to 84.43\% on REAL-Eval-Standard and from 53.95\% to 65.36\% on REAL-Eval-Fine, gains of 16.01 and 11.41 percentage points. It achieves the strongest Standard result, surpasses all evaluated open-source speech-understanding and reward models on both tiers, and ranks second overall with a 70.64\% macro-average. These results demonstrate that the complete REAL post-training pipeline improves pairwise preference prediction over the MiMo-Audio backbone. The 11.41 percentage point gain on REAL-Eval-Fine, whose reference texts, reference speech, and source TTS systems are disjoint from training, further supports generalization to the unseen systems evaluated in this benchmark. However, REAL-Judge remains 1.95 and 2.27 percentage points behind Gemini-3.1-Pro-Preview and Qwen3.5-Omni, respectively, on REAL-Eval-Fine. The remaining gap underscores the difficulty of distinguishing nuanced expressive differences across diverse high-quality speech scenarios and points to further opportunities for improving fine-grained sensitivity and cross-scenario generalization.

\subsection{Ablation Studies}

\begin{table}[b]
\vspace{-2.0em}
\centering
\caption{Time-marker ablation. Values are preference accuracy (\%).}
\label{tab:time-marker-ablation}
\begin{tabular}{lcc}
\toprule
\textbf{Model} & \textbf{Standard} & \textbf{Fine} \\
\midrule
REAL-Judge & \textbf{84.43} & \textbf{65.36} \\
\midrule
w/o time marker & 82.92 & 63.77 \\
w/ time marker (1s) & 83.00 & 64.07 \\
\bottomrule
\end{tabular}
\end{table}

Table~\ref{tab:time-marker-ablation} shows that removing time markers decreases accuracy by 1.51 and 1.59 percentage points on the two tiers, demonstrating the benefit of providing explicit temporal coordinates in the audio input for both salient defect detection and fine-grained preference discrimination. Notably, a 1-second setting yields only marginal improvements over the marker-free variant and remains below the default 2-second setting. Because markers are introduced only during post-training, overly dense markers may fragment the audio representation; 2-second granularity provides a better balance between temporal resolution and modeling complexity.

Beyond the time-marker ablation, we examine whether collaborative deliberation improves evidence construction. On REAL-Eval-Standard, REAL-Jury achieves 80.92\% accuracy, compared with 79.82\% for Qwen3.5-Omni and 75.33\% for Step-Audio-R1.1. These results support the benefit of coordinating complementary observations and revisiting disputed claims over relying on an individual evidence constructor.

\subsection{Model-Level Agreement with Human Evaluation}

\begin{figure}[t]
\centering
\includegraphics[width=0.7\columnwidth]{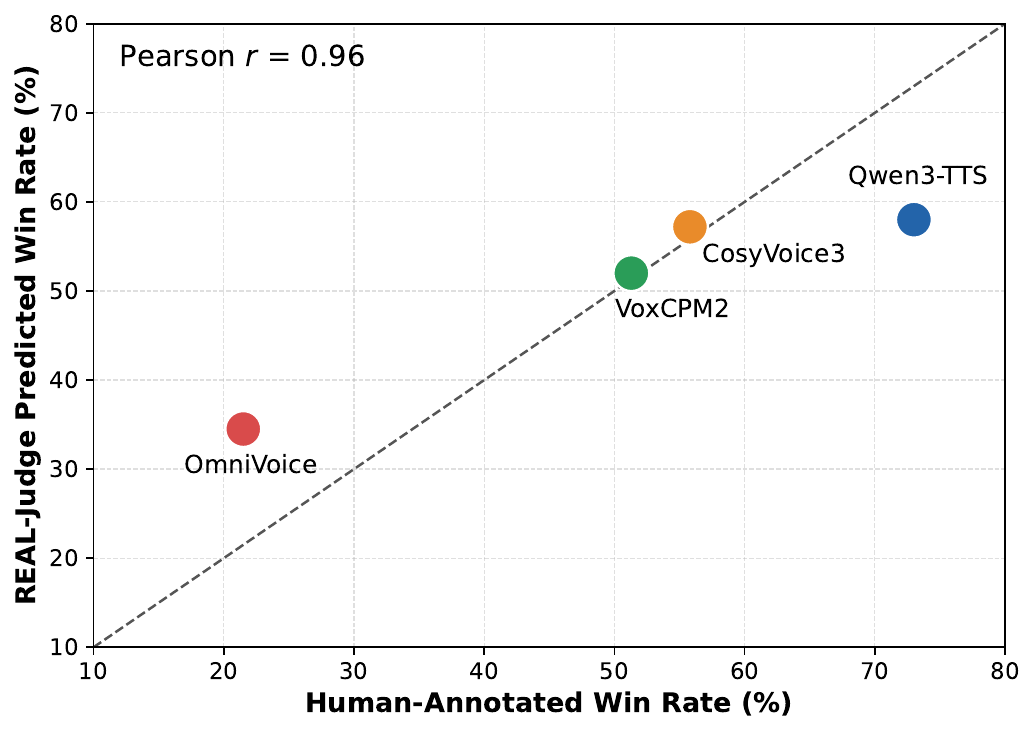}
\caption{Predicted versus human-annotated model win rates. The dashed line denotes perfect calibration.}
\label{fig:model-win-rate-correlation}
\vspace{-1.0em}
\end{figure}

Figure~\ref{fig:model-win-rate-correlation} examines whether the sample-level preference judgments of REAL-Judge can be aggregated into system-level evaluation results. The predicted and human-annotated win rates exhibit a strong positive model-level correlation, with a Pearson coefficient of $r=0.96$, and the ordering of all four systems is preserved. In particular, the predicted win rates of CosyVoice3 and VoxCPM2 are 57.4\% and 51.1\%, closely matching their human-annotated win rates of 56.7\% and 51.0\%, respectively. REAL-Judge underestimates Qwen3-TTS by 14.6 percentage points and overestimates OmniVoice by 13.4 percentage points, suggesting that its predictions are compressed toward the middle of the win-rate range. Despite these calibration errors at the extremes, the consistent system ordering suggests that sample-level judgments can support meaningful model-level comparisons. 
REAL-Judge may therefore serve as an automatic pre-screening tool and potentially reduce evaluation costs.

\section{Conclusion}

In this work, we propose REAL, a unified framework that transforms utterance-level human preferences into temporally referenced and structured supervision for generated-speech evaluation. At its core, REAL-Jury organizes independent model observations through timestamp alignment and dispute-oriented reconciliation, producing structured preference reports that REAL-Judge subsequently distills into a single timestamp-aware reward model. REAL-Eval completes the framework with a two-tier benchmark spanning salient synthesis defects and subtle perceptual differences. Across both tiers, REAL-Judge delivers substantial gains over its MiMo-Audio backbone and surpasses all evaluated open-source speech reward models. Its agreement with human judgments at the system level further suggests that evidence-linked reward modeling offers a promising path toward scalable, interpretable, and human-aligned TTS evaluation. Future work will extend the training data to English speech and explore more effective post-training strategies, including on-policy distillation (OPD) and reinforcement learning (RL), to improve model generalization.

\clearpage
\begingroup
\footnotesize
\bibliographystyle{IEEEbib}
\bibliography{refs}

@article{cosyvoice2,
  title={Cosyvoice 2: Scalable streaming speech synthesis with large language models},
  author={Du, Zhihao and Wang, Yuxuan and Chen, Qian and Shi, Xian and Lv, Xiang and Zhao, Tianyu and Gao, Zhifu and Yang, Yexin and Gao, Changfeng and Wang, Hui and others},
  journal={arXiv preprint arXiv:2412.10117},
  year={2024}
}

@inproceedings{maskgct,
  title={Maskgct: Zero-shot text-to-speech with masked generative codec transformer},
  author={Wang, Yuancheng and Zhan, Haoyue and Liu, Liwei and Zeng, Ruihong and Guo, Haotian and Zheng, Jiachen and Zhang, Qiang and Zhang, Xueyao and Zhang, Shunsi and Wu, Zhizheng},
  booktitle={International Conference on Learning Representations},
  volume={2025},
  pages={47127--47150},
  year={2025}
}

@inproceedings{F5-TTS,
  title={F5-tts: A fairytaler that fakes fluent and faithful speech with flow matching},
  author={Chen, Yushen and Niu, Zhikang and Ma, Ziyang and Deng, Keqi and Wang, Chunhui and Zhao, Jian and Yu, Kai and Chen, Xie},
  booktitle={Proceedings of the 63rd Annual Meeting of the Association for Computational Linguistics (Volume 1: Long Papers)},
  pages={6255--6271},
  year={2025}
}

@article{goat_tts,
  title={Goat-tts: Expressive and realistic speech generation via a dual-branch llm},
  author={Song, Yaodong and Chen, Hongjie and Lian, Jie and Zhang, Yuxin and Xia, Guangmin and Li, Zehan and Zhao, Genliang and Kang, Jian and Li, Jie and Li, Yongxiang and others},
  journal={arXiv preprint arXiv:2504.12339},
  year={2025}
}

@book{itu_p800,
  title={Methods for subjective determination of transmission quality},
  author={International Telecommunication Union. Telecommunication Standardization Sector},
  year={1996},
  publisher={International Telecommunication Union}
}

@article{mosnet,
  title={Mosnet: Deep learning based objective assessment for voice conversion},
  author={Lo, Chen-Chou and Fu, Szu-Wei and Huang, Wen-Chin and Wang, Xin and Yamagishi, Junichi and Tsao, Yu and Wang, Hsin-Min},
  journal={arXiv preprint arXiv:1904.08352},
  year={2019}
}

@article{utmos,
  title={Utmos: Utokyo-sarulab system for voicemos challenge 2022},
  author={Saeki, Takaaki and Xin, Detai and Nakata, Wataru and Koriyama, Tomoki and Takamichi, Shinnosuke and Saruwatari, Hiroshi},
  journal={arXiv preprint arXiv:2204.02152},
  year={2022}
}

@article{scoreq,
  title={SCOREQ: Speech quality assessment with contrastive regression},
  author={Ragano, Alessandro and Skoglund, Jan and Hines, Andrew},
  journal={Advances in Neural Information Processing Systems},
  volume={37},
  pages={105702--105729},
  year={2024}
}

@article{uppsqa,
  title={Universal preference-score-based pairwise speech quality assessment},
  author={Shi, Yu-Fei and Ai, Yang and Ling, Zhen-Hua},
  journal={arXiv preprint arXiv:2506.01455},
  year={2025}
}

@article{wavreward,
  title={Wavreward: Spoken dialogue models with generalist reward evaluators},
  author={Ji, Shengpeng and Liang, Tianle and Li, Yangzhuo and Zuo, Jialong and Fang, Minghui and He, Jinzheng and Chen, Yifu and Liu, Zhengqing and Jiang, Ziyue and Cheng, Xize and others},
  journal={arXiv preprint arXiv:2505.09558},
  year={2025}
}

@inproceedings{speechjudge,
  title={Speechjudge: Towards human-level judgment for speech naturalness},
  author={Zhang, Xueyao and Wang, Chaoren and Liao, Huan and Li, Ziniu and Wang, Yuancheng and Wang, Li and Jia, Dongya and Chen, Yuanzhe and Li, Xiulin and Chen, Zhuo and others},
  booktitle={International Conference on Learning Representations},
  volume={2026},
  pages={86261--86292},
  year={2026}
}

@inproceedings{speechllmjudges,
  title={Speechllm-as-judges: Towards general and interpretable speech quality evaluation},
  author={Wang, Hui and Zhao, Jinghua and Yang, Yifan and Liu, Shujie and Chen, Junyang and Zhang, Yanzhe and Zhao, Shiwan and Li, Jinyu and Zhou, Jiaming and Sun, Haoqin and others},
  booktitle={Proceedings of the 64th Annual Meeting of the Association for Computational Linguistics (Volume 1: Long Papers)},
  pages={7675--7700},
  year={2026}
}

@inproceedings{unisrm,
  title={UniSRM: A Unified Speech Reward Model for Reasoning-Based Fine-grained Assessment},
  author={Wang, Yuanyuan and Yang, Dongchao and Deng, Yayue and Wu, Zhiyong and Guo, Steven Y and Meng, Helen and Wu, Xixin},
  booktitle={Proceedings of the 64th Annual Meeting of the Association for Computational Linguistics (Volume 1: Long Papers)},
  pages={46346--46366},
  year={2026}
}

@article{ttsprism,
  title={TTS-PRISM: A Perceptual Reasoning and Interpretable Speech Model for Fine-Grained Diagnosis},
  author={Wang, Xi and Wang, Jie and Song, Xingchen and Song, Baijun and Xie, Jingran and Shao, Jiahe and Lin, Zijian and Wu, Di and Meng, Meng and Luan, Jian and others},
  journal={arXiv preprint arXiv:2604.22225},
  year={2026}
}

@article{Kuan2024ObjectHallucination,
  title={Understanding sounds, missing the questions: The challenge of object hallucination in large audio-language models},
  author={Kuan, Chun-Yi and Huang, Wei-Ping and Lee, Hung-yi},
  journal={arXiv preprint arXiv:2406.08402},
  year={2024}
}

@article{spotsound,
  title={SpotSound: Enhancing Large Audio-Language Models with Fine-Grained Temporal Grounding},
  author={Sun, Luoyi and Zhou, Xiao and Li, Zeqian and Zhang, Ya and Wang, Yanfeng and Xie, Weidi},
  journal={arXiv preprint arXiv:2604.13023},
  year={2026}
}

@article{Turpin2023UnfaithfulCoT,
  title={Language models don't always say what they think: Unfaithful explanations in chain-of-thought prompting},
  author={Turpin, Miles and Michael, Julian and Perez, Ethan and Bowman, Samuel},
  journal={Advances in Neural Information Processing Systems},
  volume={36},
  pages={74952--74965},
  year={2023}
}

@article{Lanham2023CoTFaithfulness,
  title={Measuring faithfulness in chain-of-thought reasoning},
  author={Lanham, Tamera and Chen, Anna and Radhakrishnan, Ansh and Steiner, Benoit and Denison, Carson and Hernandez, Danny and Li, Dustin and Durmus, Esin and Hubinger, Evan and Kernion, Jackson and others},
  journal={arXiv preprint arXiv:2307.13702},
  year={2023}
}

@article{qwen3.5_omni,
  title={Qwen3.5-omni technical report},
  author={Team, Qwen},
  journal={arXiv preprint arXiv:2604.15804},
  year={2026}
}

@article{moss_audio,
  title={Moss-audio technical report},
  author={Yang, Chen and Yu, Chufan and Chen, Hanfu and Zhu, Jie and Chen, Jingqi and Chen, Ke and Wang, Wenxuan and Wang, Yang and Jiang, Yaozhou and Jiang, Yi and others},
  journal={arXiv preprint arXiv:2606.01802},
  year={2026}
}

@article{lora,
  title={Lora: Low-rank adaptation of large language models},
  author={Hu, Edward J and Shen, Yelong and Wallis, Phillip and Allen-Zhu, Zeyuan and Li, Yuanzhi and Wang, Shean and Wang, Lu and Chen, Weizhu},
  journal={arXiv preprint arXiv:2106.09685},
  year={2021}
}

@inproceedings{intp,
  title={Advancing zero-shot text-to-speech intelligibility across diverse domains via preference alignment},
  author={Zhang, Xueyao and Wang, Yuancheng and Wang, Chaoren and Li, Ziniu and Chen, Zhuo and Wu, Zhizheng},
  booktitle={Proceedings of the 63rd Annual Meeting of the Association for Computational Linguistics (Volume 1: Long Papers)},
  pages={12251--12270},
  year={2025}
}

@article{omnivoice,
  title={Omnivoice: Towards omnilingual zero-shot text-to-speech with diffusion language models},
  author={Zhu, Han and Ye, Lingxuan and Kang, Wei and Yao, Zengwei and Guo, Liyong and Kuang, Fangjun and Han, Zhifeng and Zhuang, Weiji and Lin, Long and Povey, Daniel},
  journal={arXiv preprint arXiv:2604.00688},
  year={2026}
}

@article{qwen3_tts,
  title={Qwen3-tts technical report},
  author={Hu, Hangrui and Zhu, Xinfa and He, Ting and Guo, Dake and Zhang, Bin and Wang, Xiong and Guo, Zhifang and Jiang, Ziyue and Hao, Hongkun and Guo, Zishan and others},
  journal={arXiv preprint arXiv:2601.15621},
  year={2026}
}

@article{voxcpm2,
  title={Voxcpm2 technical report},
  author={Zhou, Yixuan and Zeng, Guoyang and Liu, Xin and Li, Xiang and Yu, Renjie and Gui, Jiancheng and Wu, Jiaheng and Wang, Ziyang and Shen, Xudong and Ye, Runchuan and others},
  journal={arXiv preprint arXiv:2606.06928},
  year={2026}
}

@article{cosyvoice3,
  title={Cosyvoice 3: Towards in-the-wild speech generation via scaling-up and post-training},
  author={Du, Zhihao and Gao, Changfeng and Wang, Yuxuan and Yu, Fan and Zhao, Tianyu and Wang, Hao and Lv, Xiang and Wang, Hui and Ni, Chongjia and Shi, Xian and others},
  journal={arXiv preprint arXiv:2505.17589},
  year={2025}
}

@article{deepseek_v4,
  title={Deepseek-v4: Towards highly efficient million-token context intelligence},
  author={Xu, Anyi and Lin, Bangcai and Xue, Bing and Wang, Bingxuan and Xu, Bingzheng and Wu, Bochao and Zhang, Bowei and Lin, Chaofan and Dong, Chen and Ling, Chenchen and others},
  journal={arXiv preprint arXiv:2606.19348},
  year={2026}
}

@article{step_audio_r1,
  title={Step-audio-r1 technical report},
  author={Tian, Fei and Zhang, Xiangyu Tony and Zhang, Yuxin and Zhang, Haoyang and Li, Yuxin and Liu, Daijiao and Deng, Yayue and Wu, Donghang and Chen, Jun and Zhao, Liang and others},
  journal={arXiv preprint arXiv:2511.15848},
  year={2025}
}

@article{mimo_audio,
  title={Mimo-audio: Audio language models are few-shot learners},
  author={Xiaomi, LCT},
  journal={CoRR, abs/2512.23808},
  year={2025}
}

@article{funasr,
  title={Funasr: A fundamental end-to-end speech recognition toolkit},
  author={Gao, Zhifu and Li, Zerui and Wang, Jiaming and Luo, Haoneng and Shi, Xian and Chen, Mengzhe and Li, Yabin and Zuo, Lingyun and Du, Zhihao and Xiao, Zhangyu and others},
  journal={arXiv preprint arXiv:2305.11013},
  year={2023}
}

@article{qwen2.5-omni,
      title={Qwen2.5-Omni Technical Report}, 
      author={Jin Xu and Zhifang Guo and Jinzheng He and Hangrui Hu and others},
      journal={arXiv preprint arXiv:2503.20215},
	  year={2025}
}

@article{qwen3-omni,
  title={Qwen3-omni technical report},
  author={Xu, Jin and Guo, Zhifang and Hu, Hangrui and Chu, Yunfei and Wang, Xiong and He, Jinzheng and Wang, Yuxuan and Shi, Xian and He, Ting and Zhu, Xinfa and others},
  journal={arXiv preprint arXiv:2509.17765},
  year={2025}
}

@misc{gemini31pro,
  author       = {{Google DeepMind}},
  title        = {Gemini 3.1 Pro Model Card},
  year         = {2026},
  howpublished = {\url{https://deepmind.google/models/model-cards/gemini-3-1-pro/}}
}
\endgroup

\end{document}